\documentclass{SciPost}
\usepackage[utf8]{inputenc}
\usepackage{graphicx}% Include figure files
\usepackage{epstopdf}
\usepackage{bbm}
\usepackage{bm}% bold math
\usepackage{verbatim}
\usepackage{amssymb}
\usepackage{listings}
\usepackage{color}
\definecolor{darkblue}{rgb}{0,0,0.6}
\usepackage{here}
\usepackage{mathtools}
\usepackage{dsfont}

\renewcommand\vec[1]{\boldsymbol{#1}}

\newcommand\fig[1]{Fig.~\ref{#1}}
\newcommand\Sec[1]{Sec.~\ref{#1}}

\newcommand\eq[1]{Eq.~(\ref{#1})}

\newcommand{\figwidth}{0.6\linewidth}

\hypersetup{
    colorlinks,
    linkcolor={red!50!black},
    citecolor={blue!50!black},
    urlcolor={blue!80!black}
}

\usepackage[bitstream-charter]{mathdesign}
\DeclareSymbolFont{usualmathcal}{OMS}{cmsy}{m}{n}
\DeclareSymbolFontAlphabet{\mathcal}{usualmathcal}

\fancypagestyle{SPstyle}{
\fancyhf{}
\lhead{\colorbox{scipostblue}{\bf \color{white} ~SciPost Physics }}
\rhead{{\bf \color{scipostdeepblue} ~Submission }}

\fancyfoot[C]{\textbf{\thepage}}
}

\begin{document}

\pagestyle{SPstyle}

\begin{center}{\Large \textbf{\color{scipostdeepblue}{
Dynamical Criticality of a Machine-learning-assisted \\ Monte Carlo algorithm for a Mean-Field Spin Glass model\\
}}}\end{center}

\begin{center}\textbf{
% Use (full) first name (+ middle name initials) + surname format.
% Separate subsequent authors by a comma, omit comma and use "and" for the last author.
% Mark the corresponding author(s) with a superscript symbol in this order
% \star, \dagger, \ddagger, \circ, \S, \P, \parallel, ...
Seiya Miyamoto\textsuperscript{1$\star$},
Masayuki Ohzeki\textsuperscript{1,2,3,4} and
Yoshihiko Nishikawa\textsuperscript{1,5}
}\end{center}

\begin{center}
{\bf 1} Graduate School of Information Sciences, Tohoku University, Sendai, Miyagi, 980-8579, Japan
\\
{\bf 2} Department of Physics, Institute of Science Tokyo, Meguro, Tokyo 152-8551, Japan
\\
{\bf 3} Research and Education Institute for Semiconductors and Informatics, Kumamoto University, Kumamoto, 860-8555, Japan
\\
{\bf 4} Sigma-i Co., Ltd., Tokyo 108-0075, Japan
\\
{\bf 5} Department of Physics, Nagoya University, Nagoya, Aichi 464-8602, Japan
%%%%%%%%%% END TODO: AFFILIATIONS
% Provide email address of corresponding author(s)
\\[\baselineskip]
$\star$ \href{mailto:email1}{\small miyamoto.seiya.q5@dc.tohoku.ac.jp}
% \,,\quad $\dagger$ \href{mailto:email2}{\small email2}
\end{center}

\section*{\color{scipostdeepblue}{Abstract}}
\textbf{\boldmath{%
We critically assess the performance of an autoregressive generative neural network model by applying it to an antiferromagnetic Ising model on a random regular graph. We train the network on equilibrium configurations in the low-temperature spin-glass phase of the model and perform Monte Carlo simulations using spin configurations generated by the network. 
The relaxation time of the Monte Carlo simulations drastically decreases with increasing the size of the training dataset and converges to an optimal value. The dynamical exponent characterizing the growth of the optimal relaxation time as a function of the system size is slightly reduced compared to the local Monte Carlo dynamics. However, we find that the size of the training dataset to achieve the optimal performance grows much faster with the system size than the relaxation time, implying that the total training cost eventually hinders a practical use of the method at large system sizes.
}}

\vspace{\baselineskip}

%%%%%%%%%% BLOCK: Copyright information
% This block will be filled during the proof stage, and finilized just before publication.
% It exists here only as a placeholder, and should not be modified by authors.
\noindent\textcolor{white!90!black}{%
\fbox{\parbox{0.975\linewidth}{%
\textcolor{white!40!black}{\begin{tabular}{lr}%
  \begin{minipage}{0.6\textwidth}%
    {\small Copyright attribution to authors. \newline
    This work is a submission to SciPost Physics. \newline
    License information to appear upon publication. \newline
    Publication information to appear upon publication.}
  \end{minipage} & \begin{minipage}{0.4\textwidth}
    {\small Received Date \newline Accepted Date \newline Published Date}%
  \end{minipage}
\end{tabular}}
}}
}
%%%%%%%%%% BLOCK: Copyright information

% For convenience during refereeing we turn on line numbers:
% \linenumbers

\vspace{10pt}
\noindent\rule{\textwidth}{1pt}
\tableofcontents
\noindent\rule{\textwidth}{1pt}
\vspace{10pt}

% Write your article contents here, starting from first \section.
% An example structure is given below.

\section{Introduction}
In complex systems with interesting physics, the local Markov-chain Monte Carlo (MCMC) algorithm often becomes sluggish, with unfeasibly long equilibration timescales. 
Designing efficient and rigorously correct algorithms is crucial to understanding such complex systems, but it is, in general, a nontrivial, challenging task. Since the late 80s, many efficient Monte Carlo algorithms with nonlocal or unphysical dynamical rules have been devised with several distinct strategies, ranging from cluster algorithms \cite{swendsen1986replica,swendsen1987nonuniversal,wolff1989collective,kandel1990cluster,dress1996cluster,houdayer2001cluster,liu2004rejection,zhu2015efficient} and extended and generalized ensemble methods \cite{berg1991multicanonical,berg1992multicanonical,marinari1992simulated,lyubartsev1992new,hukushima1996exchange} to nonreversible algorithms \cite{bernard2009event,suwa2010markov,turitsyn2011irreversible,ichiki2013violation,michel2014generalized,michel2015event,ohzeki2015langevin,nishikawa2015event,lei2019event,michel2020forward,ghimenti2024irreversible,nishikawa2025irreversible}. Whereas each algorithm can accelerate the convergence to equilibrium in some classes of systems, no algorithm can fully resolve the problem of slow equilibration in large glassy systems at very low temperature.

Machine-learning-assisted Monte Carlo (MLMC) \cite{noe2019boltzmann,mcnaughton2020boosting,gabrie2022adaptive,scriva2023accelerating,ciarella2023machine,jung2024normalizing,biazzo2024sparse,salakhutdinov2024diffusive,tamagnone2024coarse,del2025performance,del2025nearest,delbono2026demonstrating} has recently emerged as a novel approach to simulations of complex systems. In these algorithms, we have a pretrained neural-network model generate a new configuration that is statistically independent of a current one, and accept it with a probability that forces the system to converge to thermal equilibrium. The incorporated neural-network models are carefully designed so that the probability of generating any configuration can be exactly and easily computed, thereby allowing the exact computation of the acceptance probability as well. Those MLMC algorithms, if successful, can quickly decorrelate a system even at very low temperature and accelerate simulations by several orders of magnitude, as was demonstrated for some statistical-physics models \cite{mcnaughton2020boosting,biazzo2023autoregressive,biazzo2024sparse,del2025performance}.
However, their performance should strongly depend on the complexity of systems of interest and, in fact, for a class of glassy systems with a discontinuous replica symmetry breaking (RSB), some representative MLMC algorithms cannot sample efficiently and are even worse than naive local algorithms \cite{ciarella2023machine,ghio2024sampling}. This difficulty originates from the discontinuous nature of the glass transition in those systems \cite{ghio2024sampling}, suggesting an intrinsic limit of generative models. Nevertheless, it is not fully understood yet what classes of complex systems the generative models can accurately approximate, and further exploring the limit of the MLMC in other classes of systems would provide insights into the connection between physics and sampling efficiency.

In this paper, we assess the performance of an MLMC algorithm incorporating Masked Autoencoder for Distribution Estimation (MADE) \cite{pmlr-v37-germain15}, one of the state-of-the-art generative models that allow efficient computation of the acceptance probability, by applying it to an antiferromagnetic Ising model on a random regular graph. We dub this algorithm as MADEMC throughout the paper. The Ising model we consider exhibits another kind of glass transitions with full-step RSB \cite{zdeborova2010conjecture,coja2022ising} and has a long timescale in the low-temperature spin glass phase. At zero magnetic field, this model rigorously has $\mathbb Z_2$ symmetry, and the Boltzmann distribution is invariant with respect to the global spin inversion. Whereas more efficient and improved generative neural network models that take into account this sort of symmetric structure could be devised \cite{allingham2024generative,jun2020distribution,kohler2020equivariant}, we stick with one that does not assume any symmetric structure for the target distribution. Furthermore, to prevent any unintended advantage for the generative model, we explicitly break the $\mathbb Z_2$ symmetry of the model by introducing a magnetic field that remains sufficiently weak to preserve the spin glass phase at low temperatures.

We aim to understand the best possible performance of the generative model, as well as the computational costs required to achieve it. To this end, we train a MADE network for each random instance of the Ising model on an \textit{ideal} dataset, that is, a set of statistically independent configurations sampled in the spin glass phase, while systematically varying the size of the training dataset. Equilibrium configurations are generated for each random instance using the exchange Monte Carlo \cite{hukushima1996exchange}, also known as parallel tempering. We note that, in previous studies for the MLMC algorithms, training configurations at low temperatures were generated using sequential tempering (ST) \cite{mcnaughton2020boosting,ciarella2023machine,del2025performance,delbono2026demonstrating}, in which a generative model is successively trained with gradually lowering temperature. Whereas ST can reduce the burden of generating training configurations at low temperatures, the generated samples may exhibit biases depending on the specific implementation of ST \cite{ciarella2023machine}. Our setup should correspond to a limiting case of ST, where every temperature change is infinitesimally small. 
For quantitative comparisons between MADEMC and a local algorithm, we focus on an autocorrelation function for both algorithms. The speedup of MADEMC is then quantified through the relaxation timescales of the autocorrelation and their dependencies both on the size of the training dataset and on the system size. We further analyze in detail the dynamics of MADEMC for each random instance and argue the effects of {hard} instances on the typical performance of MADEMC.

The rest of this paper is as follows. The next section introduces the Metropolis--Hastings framework for MCMC and the local and the MADEMC algorithms. In Sec.~\ref{seq:Ising Model}, we introduce the Ising model and details of our simulations and MADE network. In Sec.~\ref{seq:Results}, we present our results on the static properties of the Ising model and our analysis on the dynamics of our MADEMC algorithm. The final section concludes the paper.

\section{Background}\label{seq:Methods}
\subsection{Metropolis--Hasting algorithm}
Whereas the MCMC can sample any target probability distribution, we restrict ourselves to the Boltzmann distribution for concreteness in the following discussions.
The Boltzmann distribution $\pi(\vec \sigma) = \exp(-\beta H(\vec \sigma)) / Z$ depends on the configuration $\vec \sigma$, the system Hamiltonian $H$, and inverse temperature $\beta$. The partition function $Z$ is unknown a priori and requires an exponentially large number of computations in general. Direct sampling from $\pi$ is thus challenging and impractical when $\vec \sigma$ lives in a high-dimensional space. The MCMC method elegantly circumvents the computation of $Z$ by repeatedly updating a configuration, thereby converging to $\pi$ for any initial distribution.

Most of the MCMC algorithms known so far rely on the framework of the Metropolis--Hastings (MH) algorithm, in which the transition probability $W(\vec \sigma^\prime | \vec \sigma)$ from an instantaneous configuration $\vec \sigma$ to another one $\vec \sigma^\prime$ reads
\begin{align}
    W(\vec \sigma^\prime | \vec \sigma) 
    = Q({\vec{\sigma}^\prime | \vec{\sigma}}) A(\vec{\sigma}^\prime | \vec{\sigma}).
\end{align}
Here, $Q$ and $A$ are the proposal and acceptance probabilities, respectively. For a given $Q$, the detailed balance condition yields
\begin{align}
    \label{eq:MH acceptance probability}
    A(\vec\sigma^\prime | \vec\sigma) = \min \left( 
        1, \frac{\pi(\vec \sigma^\prime)}{\pi(\vec\sigma)}\frac{Q(\vec\sigma | \vec\sigma^\prime)}{Q(\vec\sigma^\prime | \vec\sigma)}
    \right).
\end{align}
For Ising spin systems, which we will study below, one typically chooses one spin randomly and flips it in one Monte Carlo trial. This yields a symmetric $Q$, i.e. $Q({\vec{\sigma}^\prime | \vec{\sigma}}) = Q({\vec{\sigma} | \vec{\sigma}^\prime})$, and the acceptance probability reduces to the original Metropolis probability 
\begin{equation}
    A(\vec{\sigma}' | \vec{\sigma})= \min \left( 
        1, \frac{\pi(\vec{\sigma}')}{\pi(\vec{\sigma})}
    \right)
    = \min \left(
        1, e^{-\beta \Delta H}
    \right),
\end{equation}
where $\Delta H = H(\vec{\sigma^\prime}) - H(\vec{\sigma})$. Thanks to the local nature of the algorithm, it can be applied to any system with similar symmetric probabilities for $Q$. However, the local algorithm often suffers from slow relaxation and long correlation times, especially near phase transition points and in glassy systems, making it difficult to obtain statistically independent samples within feasible simulation times. Devising clever Monte Carlo moves beyond the local updates is essential to reducing the slowdown, but it is a challenging task for these complex systems.

\subsection{Machine-learning-assisted Monte Carlo with MADE}
In the MH algorithm, any proposal probability $Q$ can be incorporated and yields a rigorous algorithm, as far as $Q(\vec \sigma | \vec \sigma^\prime) / Q(\vec \sigma^\prime | \vec \sigma) < \infty$ in Eq.~\eqref{eq:MH acceptance probability}. Recent studies have attempted to construct a proposal probability that eliminates the slow dynamics by leveraging the power of modern generative neural networks (NNs) \cite{noe2019boltzmann,mcnaughton2020boosting,gabrie2022adaptive,scriva2023accelerating,ciarella2023machine,jung2024normalizing}. These MLMC methods work with pretrained NNs that generate a new configuration $\vec \sigma$ independent of the current one and allow direct and exact calculation of the probability of generating $\vec \sigma$. We denote this probability by $P_{\vec \theta}(\vec \sigma)$, where $\vec \theta$ indicates a set of parameters of a generative NN. The parameters $\vec \theta$ are optimized on a training dataset so that $P_{\vec \theta}(\vec \sigma)$ approximates $\pi(\vec \sigma)$ well. Ideally, if $P_{\vec \theta}(\vec \sigma)$ is exactly equal to $\pi(\vec \sigma)$, the acceptance probability 
\begin{equation}
    A(\vec{\sigma}^\prime | \vec{\sigma}) = \min \left( 
        1, \frac{\pi(\vec{\sigma}^\prime)}{\pi(\vec{\sigma})}\frac{P_{\vec{\theta}}(\vec{\sigma})}{P_{\vec{\theta}}(\vec{\sigma}^\prime)}
    \right),
    \label{eq: MH prob for generative models}
\end{equation}
is equal to one, realizing a rejection-free, direct-sampling algorithm. In reality, however, achieving an exact match between the two distributions is unattainable, and the performance of the MLMC algorithms strongly depends on the size and quality of the training dataset, the architecture of NNs, optimization algorithms for $\vec \theta$ etc.

Autoregressive generative NNs, including MADE, which we study in this paper, provide an analytically tractable representation for the probability $P_{\vec \theta}$, and thus can be easily incorporated with the MH algorithm. More precisely, $P_{\vec \theta}(\vec \sigma)$ can be decomposed as
\begin{align}
    \label{eq:autoregressive probability}
    P_{\vec{\theta}}(\vec{\sigma})
    &= \prod_{i=1}^N p^{(i)}_{\vec{\theta}}(\sigma_i | \vec{\sigma}_{<i}),
\end{align}
where $p_{\vec \theta}^{(i)}(\cdot | \vec \sigma_{<i})$ is the conditional probability for a given $\vec \sigma_{<i} = (\sigma_1, \sigma_2, \cdots, \sigma_{i-1}, 0, \cdots, 0)$. These NN models also allow us to generate a configuration according to $P_{\vec \theta}$ using the ancestral sampling technique \cite{bishop2006pattern}: We first sample $\sigma_1$ from the distribution $p^{(1)}_{\vec \theta}$, and then determine $\sigma_2$ according to the conditional probability $p^{(2)}_{\vec \theta}(\cdot | \sigma_1)$. The $i$-th component $\sigma_i$ is sampled in the same manner, i.e., $\sigma_i \sim p^{(i)}_{\vec \theta}(\cdot | \vec \sigma_{<i})$.

\begin{figure}[t]
    \centering
    \includegraphics[width=\figwidth]{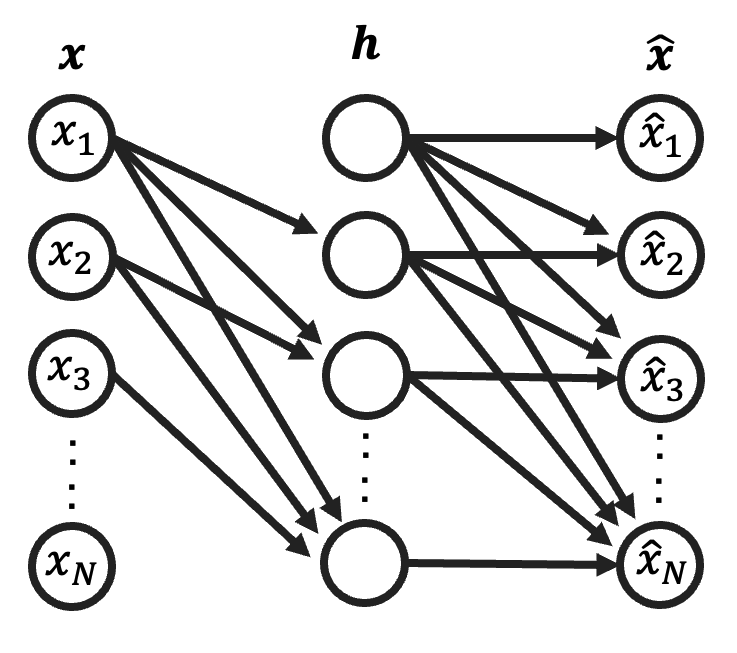}
    \caption{Example of MADE network architecture with the autoregressive property, with input $\vec{x}$, one hidden layer $\vec{h}$, and output $\hat{\vec x}$.}
    \label{fig:MADE architecture}
\end{figure}

MADE \cite{pmlr-v37-germain15} is an autoencoder NN model with its architecture having some masked connections between layers, which yield the autoregressive property, in contrast to conventional, fully-connected neural networks \cite{Hinton2006reducing}, see \fig{fig:MADE architecture}. 
In the following, we assume that input vectors are binary, i.e., each vector component is either $0$ or $1$, which suffices for this study considering an Ising system. 
For a given input vector $\vec x$, a MADE outputs a set of the probabilities $\hat{\vec x} = (p^{(i)}_{\vec \theta}(x_i = 1 | \vec x_{<i}))_i$, from which we can easily compute $P_{\vec \theta}(\vec x)$ through \eq{eq:autoregressive probability} as
\begin{equation}
    \log P_{\vec \theta}(\vec x) = \sum_i x_i \log \hat{x_i} + (1 - x_i) \log(1 - \hat{x_i}).
    \label{eq: cross entropy}
\end{equation}
Although \textit{shallow} MADE with no hidden layer can precisely approximate the Boltzmann distribution of Ising spin models \cite{ciarella2023machine,del2025performance}, we consider a more general and expressive architecture with one hidden layer here. The hidden and output layers, $\vec h$ and $\hat{\vec x}$ respectively, for such networks are explicitly given as
\begin{align}
    \begin{split}
        \vec{h} &= \phi((M^{(W)} \odot W)\vec{x} + \vec{b}),\\
        \hat{\vec x} &= S((M^{(V)} \odot V)\vec{h} + \vec{c}),   
    \end{split}
    \label{eq: MADE output}
\end{align}
where $\phi(\cdot)$ and $S(\cdot)$ are activation and the standard sigmoid functions, respectively, and the operator $\odot$ denotes the Hadamard product. The weight matrices $W$ and $V$, the mask matrices $M^{(W)}$ and $M^{(V)}$, and the bias vectors $\vec b$ and $\vec c$ specify a MADE network. The mask matrices are given and fixed so as to have the network satisfy the autoregressive property \cite{pmlr-v37-germain15}, while the others are optimized so that $P_{\vec \theta}$ well approximates the Boltzmann distribution. See \Sec{sec: details on MADE} for more details on the optimization and our setup for the mask matrices.

When running an MLMC simulation with MADE, we generate a configuration from $P_{\vec \theta}$ and accept it with the probability of \eq{eq: MH prob for generative models}. To be more specific about the generation process, we first sample $x_1$ from $p_{\vec \theta}^{(1)}$, which is given in \eq{eq: MADE output}, by feeding an arbitrary vector into a MADE network. We next input a vector whose first component is $x_1$ and obtain $p_{\vec \theta}^{(2)}(\cdot | x_1)$, from which $x_2$ is sampled. Similarly, $x_3$ is sampled from $p_{\vec \theta}^{(3)}(\cdot | x_1, x_2)$ obtained by feeding the network an input with $x_1$ and $x_2$ as its first two components. Repeating this procedure sequentially until $x_N$ is sampled completes the generation of a configuration $\vec x$.

\section{Model and Method}
\label{seq:Ising Model}

\subsection{Model}
We apply the MADEMC algorithm to an antiferromagnetic Ising model on a random $3$-regular graph with $N$ spins. The Hamiltonian of the model is
\begin{equation}
    \label{eq:hamiltonian}
    H_{\vec J}(\vec{\sigma}) = \sum_{i < j} J_{ij} \sigma_i \sigma_j - h \sum_i \sigma_i,
\end{equation}
where $J_{ij}$ and $h$ denote the coupling constant and an external magnetic field, respectively. We set $J_{ij} = J > 0$ for pairs of connected spins on a random graph and $J_{ij} = 0$ otherwise and $h = J/4$. In the context of combinatorial optimization, the ground states of this model with $h = 0$ correspond to maximum cuts (MaxCuts) of a graph, whereas, for $h \in (1, 3]$, they match to maximum independent sets (MISs) \cite{takahashi2019phase}. The worst-case computational complexity of finding a MaxCut and an MIS belongs to the class \textsf{NP-hard}. Nevertheless, the average-case complexity should depend on an ensemble of random graphs. This was explicitly shown for the MIS problem on Erd\H{o}s-R\'enyi (ER) graphs using an exact algorithm in Ref.~\cite{takahashi2017exact}: When the average connectivity $c$ is small enough, the computational time increases polynomially with $N$, while, for a larger $c$, it grows exponentially. This sort of \textit{easy-to-hard} transition often accompanies a zero-temperature spin-glass transition with replica symmetry breaking (RSB) in many optimization problems \cite{cocco2001statistical,zhou2003vertex,achlioptas2008algorithmic}. For random $d$-regular graphs with $d \geq 3$, the spin glass phase was rigorously proved to be stable at finite low temperature when $h = 0$ \cite{coja2022ising}, which should also survive with a small enough magnetic field, as in other mean-field spin glass models \cite{jorg2008behavior}. We will numerically show that a spin glass transition indeed takes place at finite temperature in the model with our choice $h = J/4$.

\subsection{Details of Monte Carlo simulation and generation of training data}

We employ the exchange Monte Carlo method \cite{hukushima1996exchange}, also known as parallel tempering, for the study of static properties of the system and generating training configurations. We prepare $M$ replicas for a temperature sequence $T_1, T_2, \cdots T_M$ satisfying $T_m < T_{m+1}$ ($m=1,\cdots,M-1$), where $T_i$ is the temperature of the $i$-th replica. The temperatures are set so as to have the ratio of adjacent temperatures $T_{m+1}/T_m$ constant. The minimum and maximum temperatures are $T_1=1$ and $T_M=2$. Temperature swaps are performed between adjacent temperatures every Monte Carlo sweep (MCS) per spin. 
The number of replicas $M$ is set so that the swap probability is sufficiently high and the dynamics in the temperature space is fast enough. Based on preliminary runs on representative random graph instances, we set $M = 16$ for $N \leq 256$ and $M = 32$ for $N \geq 512$. 
For each system size, we present simulation results averaged over $1024$ random graphs in the following.

We recall that our aim in this paper is to test the performance of optimized MADE for a hard problem. We thus train a MADE network on a set of equilibrium, independent spin configurations generated at $T / J = 1$, well below the spin glass transition temperature of the model $T_c / J \approx 1.13$ (see \Sec{seq:Results}). A spin configuration is sampled every $\tau_\text{output}$ sweeps per spin, where $\tau_\text{output}$ is set to sufficiently large for pairs of two successive output configurations to have a very small correlation $\lesssim 0.1$. In our simulations, $\tau_\text{output}$ ranges from $10$ to $40$, depending on the system size.

\subsection{Architecture and training of MADE network}
\label{sec: details on MADE}

The set of parameters for MADE $\vec\theta=\{\vec{W}, \vec{V}, \vec{b}, \vec{c}\}$ is optimized so as to minimize the Kullback--Leibler (KL) divergence between $P_{\vec \theta}$ and a target distribution $\pi$
\begin{equation}
    D_\text{KL}[\pi \mid \mid P_{\vec{\theta}}] = \sum_{\vec x} \pi(\vec{x}) \log \frac{\pi(\vec{x})}{P_{\vec{\theta}}(\vec{x})},
\end{equation}
from which the optimal parameter set $\vec{\theta}^*$ is given as
\begin{align}
    \vec{\theta}^* &= \underset{\vec{\theta}} {\operatorname{argmin}} \,D_\text{KL}[\pi \mid \mid P_{\vec{\theta}}] \notag \approx \underset{\vec{\theta}} {\operatorname{argmin}} \,\left[-\sum_{d=1}^D \log P_{\vec{\theta}}(\vec{x}^{(d)})\right] 
\end{align}
where $\{\vec x^{(d)}\}_{d=1}^D$ is a training set of configurations following the target distribution $\pi$.

We use a MADE network with one hidden layer for all of the system sizes, which is expressive enough to precisely approximate the Boltzmann distribution of some spin systems \cite{ciarella2023machine,del2025performance}. The dimension of the hidden layer $N_h$ is set to $4N$. We have checked that no significant improvement is found in performance by further increasing the hidden layer dimension.

As MADE generates spin values sequentially from spin $1$ to $N$ and a value for spin $i$ is sampled from a distribution conditioned on $\vec \sigma_{<i}$, the labeling of spin indices should affect the performance. We label spins so that nearby spins have similar indices to take into account correlations between interacting spins. To be more specific, starting from a randomly chosen spin, we assign indices to spins in breadth-first search order.
Each unit in the hidden layer is also assigned an index $m(k) = (k - 1) \bmod N$ $(k=1, \cdots, N_h)$, which specifies the mask matrices $M^{(W)}$ and $M^{(V)}$ as
\begin{align}
    \begin{split}
        M^{(W)}_{ik} &= \mathbf{1}\!\left( m(k) \geq i \right) \\
        M^{(V)}_{ki} &= \mathbf{1}\!\left( i > m(k) \right),        
    \end{split}
\end{align}
with $\mathbf{1}\!\left( \cdot \right)$ the indicator function.

Training of MADE is done with batch size $128$ for $100$ epochs at maximum, using RAdam Schedulefree optimizer \cite{defazio2024road, liu2019variance}. We stop the training when the validation loss does not improve for $10$ consecutive epochs. The learning rate $\eta$ of the optimizer affects the performance of a trained MADE network and should be carefully chosen. To this end, we first train the MADE network for $16$ randomly chosen instances of the Ising model for each system size, with various learning rates $\eta$ ranging from $10^{-4}$ to $10^{-1}$. We then adopt the rate that yields the shortest mean relaxation time for subsequent MADEMC simulations. See \Sec{sec: dynamics} for the definition of the relaxation time.

\section{Results}
\label{seq:Results}

\subsection{Spin glass transition}

\begin{figure}[t]
  \centering
  \includegraphics[width=\figwidth]{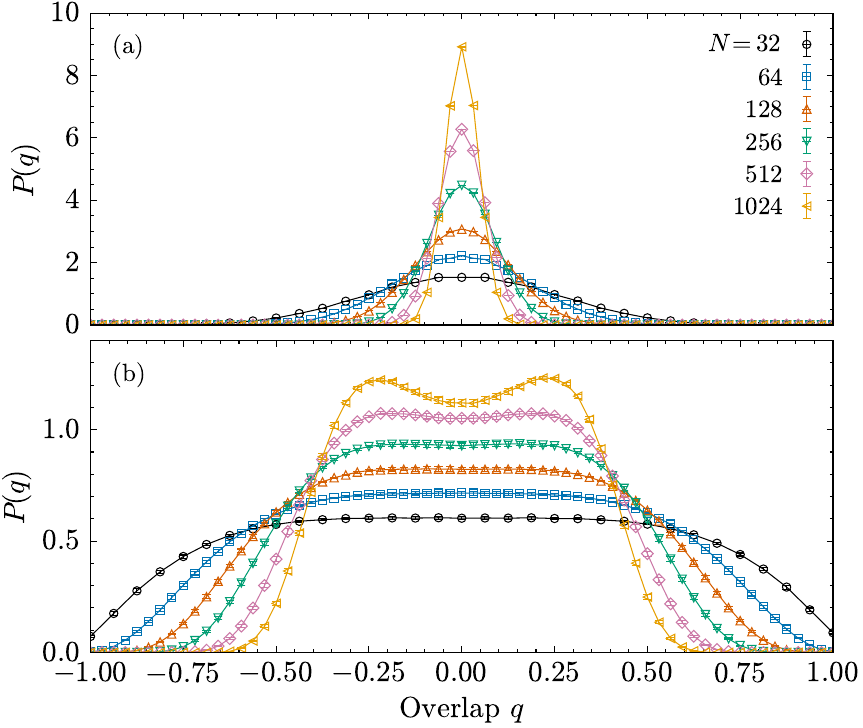}
  \caption{
    Overlap distribution $P(q)$ at temperatures (a) $T / J = 2$ and (b) $T / J = 1$. The curves are shown from bottom to top for $N=32, 64, 128, 256, 512$, and $1024$.
  }
  \label{fig:overlap distribution}
\end{figure}

To justify our choice of temperature for benchmarking our MADE network, we first discuss the spin-glass transition of the model, using results from the local Metropolis algorithm and the exchange Monte Carlo method. The antiferromagnetic model on a random $3$-regular graph with $h = 0$ has a spin glass transition at $T / J \approx 1.135$ \cite{coja2022ising}. Applying a homogeneous magnetic field should yield a slightly lower transition temperature \cite{jorg2008behavior}. We precisely estimate the critical temperature by analyzing the overlap distribution and the spin glass susceptibility.

We start our discussion with the overlap, an order parameter for the spin glass transition, and its distribution. The overlap is defined as 
\begin{align}
    q_{12} = \frac{1}{N} \sum_{i=1}^N (\sigma^{(1)}_i - m^{(1)}) (\sigma^{(2)}_i - m^{(2)}),
    \label{eq: overlap}
\end{align}
where $\vec \sigma^{(1)}$ and $\vec \sigma^{(2)}$ are two statistically independent configurations sampled from equilibrium, and $m^{(k)} = \sum \sigma_i^{(k)}/N$ is an instantaneous magnetization of configuration $k \in \{1, 2\}$. For Ising spin glass models with global $\mathbb Z_2$ symmetry, it suffices to assume $m^{(k)}$ is zero. Our model, by contrast, lacks this symmetry, and subtracting an instantaneous magnetization, as in \eq{eq: overlap}, is needed to remove trivial overlaps due to finite magnetizations. 
The overlap distribution is then defined as:
\begin{equation}
    P(q) = \left[ \langle \delta ( q - q_{12}) \rangle \right]_{\vec J},
\end{equation}
where the brackets $\langle \cdot \rangle$ and $[\cdot]_{\vec J}$ denote a thermal average and a disorder average over many random graphs, respectively.
We show in Fig.~\ref{fig:overlap distribution} $P(q)$ at $T / J = 1$ and $2$, for system sizes $N=32, 64, 128, 256, 512$, and $1024$. At $T / J = 2$, $P(q)$ exhibits a Gaussian-like shape centered at zero, with its width rapidly shrinking with $N$. This $N$ dependence implies that $P(q) \to \delta(0)$ at $N \to \infty$ and the system is in the paramagnetic phase at this temperature. At the lower temperature, $T / J = 1$, on the other hand, $P(q)$ is much broader and exhibits a nontrivial structure; at the largest $N$, $P(q)$ develops a clear double-peaked structure and does not approach a simple delta function. We thus expect a spin-glass transition at $ T_c/J> 1$. Note that similar overlap distributions have been observed in simulations of other spin glass models \cite{parisi2012numerical,marinari1998critical,parisi1999continuous}.

\begin{figure}[t]
    \centering
    \includegraphics[width=\figwidth]{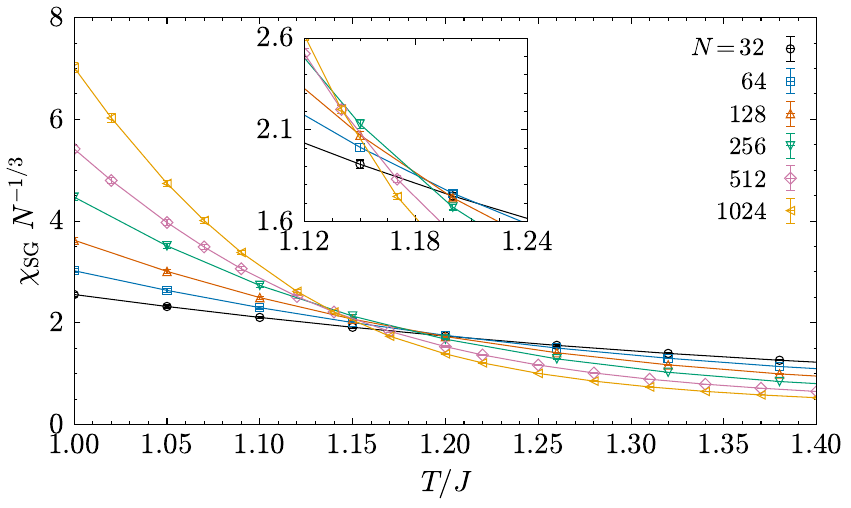}
    \caption{Scaled spin glass susceptibility $\chi_\text{SG}/N^{1/3}$. The curves for different system sizes intersect at $T/J \approx 1.13$.}
    \label{fig:spin-glass susceptibility}
\end{figure}

To precisely quantify the complex structure and the $N$ dependence of $P(q)$, we compute the spin glass susceptibility $\chi_\text{SG}$:
\begin{equation}
    \chi_\text{SG} = N \big[\langle {q_{12}}^2 \rangle - \langle {q_{12}} \rangle^2 \big]_{\vec J}.
\end{equation}
For mean-field spin glass models, $\chi_\text{SG}$ obeys the following finite-size scaling form \cite{jorg2008behavior,billoire2003numerical}:
\begin{equation}
    \chi_\text{SG}(N, T) = N^{1/3} F \big[ (T-T_c)N^{1/3} \big]
\end{equation}
where $F[\cdot]$ is a universal scaling function. This implies that the scaled susceptibility $\chi_\text{SG} / N^{1/3}$ is independent of system size $N$ at the critical temperature $T_c$. Fig.~\ref{fig:spin-glass susceptibility} shows the spin glass susceptibility $\chi_\text{SG}$ divided by $N^{1/3}$ for various system sizes. This scaled susceptibility of the two largest sizes has a clear intersection at $T / J \approx 1.13$, with its position having a weak dependence on $N$. From this plot, we conclude that the critical temperature of the model is $T_c / J \approx 1.13$. Our estimate is consistent with the critical temperature $T_c / J \approx 1.135$ in the case of $h = 0$ \cite{coja2022ising}.

\subsection{Dynamics}
\label{sec: dynamics}
\begin{figure}[t]
    \centering
    \includegraphics[width=\figwidth]{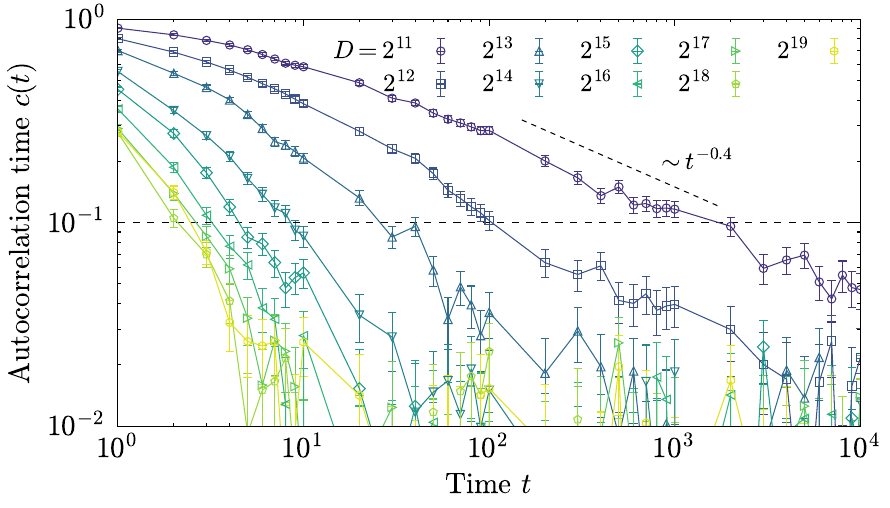}
    \caption{Autocorrelation function $c(t)$ for $N=128$ at $T=1$ and at $D = 2^{11},2^{12},...,2^{19}$. At $D \gtrsim 2^{15}$, $c(t)$ decays rapidly with an exponential form, whereas, at $D \lesssim 2^{13}$, it has a clear power-law decay $c(t) \sim t^{-\delta}$ on a long timescale. The exponent $\delta \approx 0.4$ at $D = 2^{11}$.}
    \label{fig:autocorrelation}
\end{figure}

We now turn to the dynamics of our MADEMC algorithm in the spin glass phase. Regarding the results of our static analysis of the model, we will focus on $ T/J = 1$, which is below the critical temperature. To characterize the average dynamics, we measure the autocorrelation function 
\begin{equation}
    c(t) = \frac{
        \sum_{i=1}^N \left[\langle \sigma_i(0)\sigma_i(t) \rangle - \langle \sigma_i \rangle^2 
        \right]_{\vec J}
    }{
        \sum_{i=1}^N \left[1-\langle \sigma_i \rangle^2 \right]_{\vec J}
    }, 
\end{equation}
where $\sigma_i(t)$ denotes spin $i$ at time $t$, and estimate the relaxation time $\tau$ as the number of Monte Carlo sweeps required for the autocorrelation to decay to a preset threshold value $c_\text{th}$. Here, we set $c_\text{th}$ to $10^{-1}$ and the unit of time for MADEMC to one Monte Carlo trial of a new generated configuration. The autocorrelation function and the relaxation time strongly depend on algorithms and serve as key metrics in comparing the performance of different algorithms. We quantify the speedup of the MADEMC algorithm by comparing its relaxation times with those of the local Metropolis algorithm.

\begin{figure}[t]
    \centering
    \includegraphics[width=\figwidth]{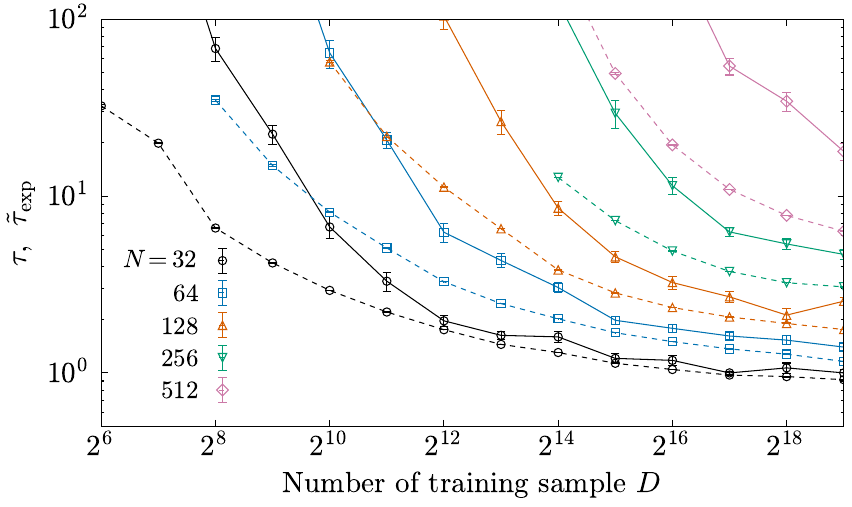}
    \caption{Relaxation time $\tau$ (solid lines) and mean time scale determined by the acceptance ratio $\tilde{\tau}_\text{exp}(D)$ (dashed lines) as functions of training data size $D$. }
    \label{fig:relaxation time}
\end{figure}

In Fig.~\ref{fig:autocorrelation}, we show $c(t)$ for $N = 128$, with different sizes of training dataset $D$. When increasing $D$, $c(t)$ decays more rapidly, consistent with a naive expectation that the approximation accuracy of the MADE network monotonically improves with $D$. The relaxation time $\tau(D, N)$ also monotonically decreases with $D$, see \fig{fig:relaxation time}. However, increasing $D$ does not keep accelerating the dynamics; instead, $\tau(D, N)$ converges to a finite value at large $D$ with its value depending on $N$. We consider the limiting value of $\tau(D, N)$ at large $D$ as the intrinsic relaxation time $\tau = \lim_{D\to\infty}\tau(D, N)$ of our MADEMC algorithm. In practice, for each $N$, we take the value of $\tau(D = 2^{19}, N)$ as $\tau$ for $N \le 256$, where the dependence of $\tau(D, N)$ on $D$ is weak. For $N = 512$, on the other hand, the relaxation time still strongly depends on $D$ up to $D = 2^{19}$, and its limiting value at $D \to \infty$ will not be included in the following analysis.

\begin{figure}[t]
    \centering
    \includegraphics[width=\figwidth]{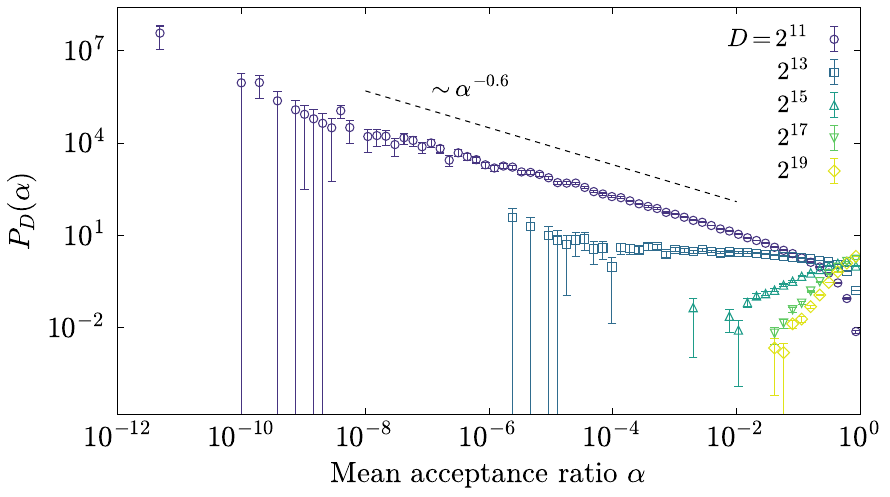}
    \caption{Distribution of the mean acceptance ratio $P_D(\alpha)$ for $D = 2^{11}, \cdots, 2^{19}$. The number of spins $N = 128$. At $D \lesssim 2^{13}$, $P_D(\alpha)$ has a clear power-law behavior, $P_D(\alpha) \sim \alpha^{-\kappa}$, with a $D$-dependent exponent $\kappa$. At $D = 2^{11}$, $\kappa \approx 0.6$.}
    \label{fig: dist of alpha}
\end{figure}

Because MADE generates a new configuration independently of the current one, the autocorrelation of each trajectory immediately goes to zero on average once a new configuration is accepted. We have verified this by confirming that $c(t)$ and the fraction of trajectories that have not transitioned yet until time $t$ match to each other with very high precision at any $D$ (not shown). The autocorrelation of each graph instance, specified by $\vec J$, thus decays exponentially with time as $\sim (1 - \alpha(D, \vec J))^t$. Here, $\alpha(D, \vec J)$ represents the mean acceptance ratio for graph instance $\vec J$ at training dataset size $D$. This exponential form then yields a timescale for the autocorrelation to decay to $c_\text{th}$,
\begin{align}
    \tau_\text{exp}(D, \vec J) = \frac{\log c_\text{th}}{\log(1 - \alpha(D, \vec J))}.
\end{align}
Naively, one would expect that the typical, average-case timescale is controlled by the mean acceptance ratio averaged over random instances, $[\alpha(D, \vec J)]_{\vec J}$, as
\begin{align}
    \tilde{\tau}_\text{exp}(D) = \frac{\log c_\text{th}}{\log(1 - [\alpha(D, \vec J)]_{\vec J})},
\end{align}
and should be close to the true relaxation time $\tau(D, N)$, which indeed holds for small $N \lesssim 64$ and large $D \gtrsim 2^{18}$. For large $N$ and small $D$, on the other hand, we notice $\tilde{\tau}_\text{exp}(D)$ and $\tau(D, N)$ are largely separated, sometimes even by a few orders of magnitude. For these parameter sets, the decay of $c(t)$ is not even exponential but algebraic; see \fig{fig:autocorrelation} for the case with $N = 128$ and $D = 2^{11}$, as an example. 

To better understand these behaviors, we compute the distribution of $\alpha(D, \vec J)$ over $\vec J$, $P_D(\alpha) = [\delta(\alpha - \alpha(D, \vec J))]_{\vec J}$. In \fig{fig: dist of alpha}, we show $P_D(\alpha)$ for various values of $D$. When $D \lesssim 2^{13}$, where the decay of $c(t)$ is algebraic and $\tau(D, N) \gg \tilde{\tau}_\text{exp}(D)$, $P_D(\alpha) \sim \alpha^{-\kappa}$ at small $\alpha$. At $D = 2^{11}$, for example, the exponent $\kappa \approx 0.6$. From the fact that decorrelation instantly takes place with single one acceptance of a new configuration, we can compute the asymptotic decay of the autocorrelation as
\begin{align}
    \begin{split}
    c(t) &= \int_0^1 d\alpha P_D(\alpha) (1 - \alpha)^t \approx \int_0^1 d\alpha \alpha^{-\kappa} (1 - \alpha)^t \sim t^{-(\kappa - 1)}.
    \end{split}
\end{align}
This is fully consistent with the asymptotic power-law decay $c(t) \sim t^{-0.4}$ at $D = 2^{11}$, see \fig{fig:autocorrelation}. These results imply that the accuracy of MADE strongly depends on $\vec J$ and that the typical performance is dominated by these \textit{hard-to-learn} instances, although they are quite rare (at $D = 2^{11}$, for example, only $5\%$ of random instances have $\alpha < 10^{-4}$).

\begin{figure}[t]
    \centering
    \includegraphics[width=\figwidth]{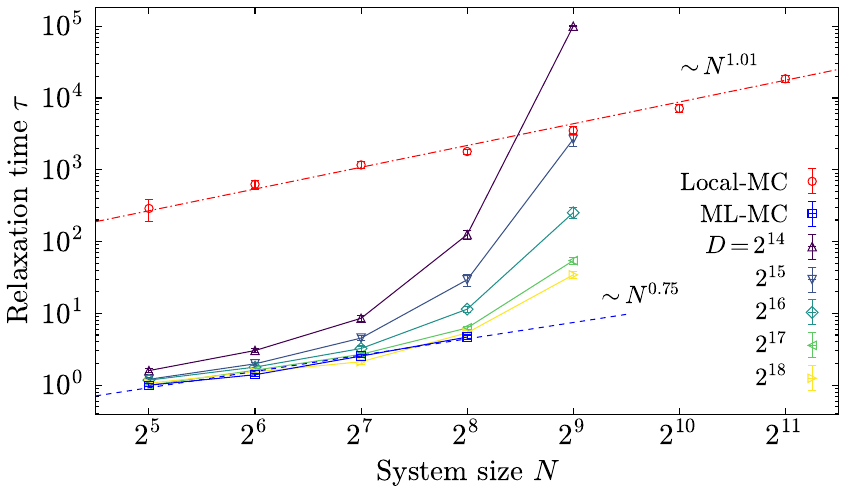}
    \caption{System-size dependence of the relaxation time $\tau$ for the local algorithm and MADEMC with various training sample sizes $D$. Blue dashed and red dash-dotted lines represent power-law curves $N^z$ with exponent $z = 1.01$ and $0.75$, respectively. }
    \label{fig:dynamic critical exponent}
\end{figure}

At large enough $D$, the power-law regime in $P_D(\alpha)$ disappears and $\tau(D, N)$ finally becomes comparable with $\tilde{\tau}_\text{exp}(D)$. The decay of $c(t)$ is exponential as expected from $P_D(\alpha)$. With increasing the system size $N$, properly training a MADE network needs more samples, and the size of the dataset, $D_c$, required to achieve the above features grows as well. We thus expect that fixing $D$ irrespective of $N$ significantly impairs the performance of MADEMC at large $N$. Indeed, the relaxation time $\tau(N, D)$ grows exponentially with $N$ when $D$ is fixed, see \fig{fig:dynamic critical exponent}. When the sample size $D$ becomes comparable with or larger than $D_c$ for each $N$, the relaxation time scales in a power-law manner with $N$, i.e., $\tau \sim N^z$. The MADEMC algorithm yields the dynamical exponent $z \approx 0.75$, whereas the local Metropolis algorithm yields a slightly larger exponent, $z \approx 1.0$. Furthermore, our algorithm accelerates the dynamics by more than two orders of magnitude. This is in stark contrast to the case where MADEMC is applied to spin models with a random first-order transition, e.g., Potts glass models \cite{ciarella2023machine,ghio2024sampling}: MADEMC fails badly and is very slow even when compared to the simple local algorithm there. Our results indicate that the free-energy structure of models with a continuous spin glass transition is much easier for the autoregressive model to approximate.

\begin{figure}[t]
    \centering
    \includegraphics[width=\figwidth]{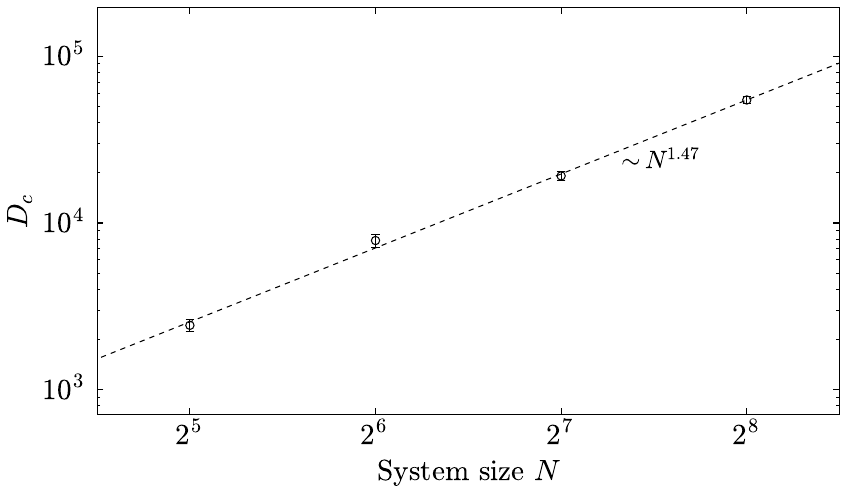}
    \caption{System-size dependence of $D_c$. Dashed line represents a power-law curve $N^\zeta$ with $\zeta = 1.47$. } 
    \label{fig:D_c  exponent}
\end{figure}

As we mentioned earlier, the sample size $D_c$ required to achieve optimal performance increases with $N$. Here, we define $D_c$ as the training sample size that is needed to have a normalized relaxation time
\begin{equation}
    \tilde \tau(D, N) = \frac{\tau(D, N) - \tau(D \to \infty, N)}{\tau(D \to \infty, N)}
    \label{eq:tau_prime}
\end{equation}
reach 2. Again, we take $\tau(D = 2^{19}, N)$ as $\tau(D \to\infty, N)$ here. In Fig.~\ref{fig:D_c  exponent}, we plot $D_c$ as a function of $N$, where $D_c$ grows algebraically with $N$ as $D_c \sim N^\zeta$, with exponent $\zeta\approx 1.5$. This exponent is significantly larger than not only the dynamical exponent $z$ for MADEMC but also the one for the local Metropolis algorithm. At large $N$, the computational costs for producing training datasets will eventually dominate the total costs and require a much longer time than naively running a local Monte Carlo simulation. Despite the significant speedup in Monte Carlo simulations, we conclude that MADEMC faces severe challenges and is impractical when applied to larger systems, which are of greatest interest to us.

\section{Conclusions}
\label{seq:Conclusions}
In summary, we have benchmarked MADE, an autoregressive neural network model, by applying it to an antiferromagnetic Ising model on a random regular graph. The Ising model exhibits a continuous spin-glass transition with a full-step RSB-like $P(q)$ and belongs to a class of hard statistical problems. We have shown that MADEMC can accelerate the dynamics of the system in the spin glass phase by a few orders of magnitude. This speedup is not merely a constant-factor gain; the MADEMC has the dynamical critical exponent $z\approx 0.75$ smaller than that for the local algorithm, meaning that the speedup is infinitely large at $N \to \infty$. While these features are definitely impressive, we have found that the number of training samples $D_c$ for achieving this optimal performance scales as $D_c \sim N^\zeta$ with $\zeta \approx 1.47$, larger than the dynamical exponent $z$. This power-law increase in training sample suggests that, when applying MADEMC to large systems, the computational cost of generating a training dataset needed for efficient simulation will be the primary bottleneck. Nevertheless, considering that the typical computational complexity of finding exact ground states of the model grows exponentially, it is surprising that only polynomially many samples are needed for efficient sampling at finite but low temperature in the spin glass phase. This could be related to the fact that for some mean-field spin glass models, an approximate ground state can be found in polynomial time \cite{montanari2021optimization}.

When $D \ll D_c$, the distribution of mean acceptance ratio $P_D(\alpha)$ has a power-law behavior toward $\alpha \to 0$, which leads to a power-law decay in the autocorrelation $c(t)$. The relaxation time at such $D$ is predominantly controlled by rare hard-to-learn instances that have a very small acceptance ratio. This effect becomes increasingly severe with the system size $N$, leading to an exponential growth of the relaxation time as a function of $N$ at fixed $D$. One would expect that the performance of MADEMC for these instances could directly correlate with their free-energy landscape structures and the shape of the overlap distribution, as is indeed true for the exchange Monte Carlo method \cite{yucesoy2013correlation}. However, we do not find any clear correlation between the acceptance ratio and the overlap distribution for each disorder instance.

We remind that, to elucidate the intrinsic ability of MADE, we have trained a MADE network with independent configurations sampled at the target temperature. In practice, however, we need to resort to more heuristic methods to generate training samples, such as sequential tempering \cite{mcnaughton2020boosting,ciarella2023machine,del2025performance,delbono2026demonstrating}. Generated training samples with sequential tempering should be more dependent and biased, resulting in worse accuracy for MADE. The scaling exponents $z$ for the relaxation time and $\zeta$ for $D_c$ that we estimated in this study are nonuniversal and would increase in more practical setups. The role of statistical correlations and biases in training samples needs to be clarified to address the scaling behaviors in realistic situations.

To design better, more expressive machine-learning models for MLMC simulations, understanding the limits of current state-of-the-art neural network models is crucial. Although the inherent limit of autoregressive and flow-based models on mean-field glassy systems is now well understood \cite{ciarella2023machine,ghio2024sampling}, their performance beyond mean field is largely unexplored. For finite-dimensional models, incorporating the spatial structure of the system into a machine-learning model should be essential to improve the accuracy. Whereas designing such architectures itself is a nontrivial task for general models \cite{biazzo2024sparse,del2025nearest}, we believe it would be fruitful to pursue this direction in future research.

\section*{Acknowledgements}
% Acknowledgements should follow immediately after the conclusion.

% % TODO: include author contributions
% \paragraph{Author contributions}
% This is optional. If desired, contributions should be succinctly described in a single short paragraph, using author initials.

% % TODO: include funding information
% \paragraph{Funding information}
% Authors are required to provide funding information, including relevant agencies and grant numbers with linked author's initials. Correctly-provided data will be linked to funders listed in the \href{https://www.crossref.org/services/funder-registry/}{\sf Fundref registry}.
This work was supported by JSPS KAKENHI Grant Nos.~22K13968, 23H01432, and 26K14992. Our study receives financial support from the Cross-Ministerial Strategic Innovation Promotion Program (SIP) from the Cabinet Office.

\bibliography{references.bib}

%%%%%%%%%% END TODO: BIBLIOGRAPHY

\end{document}